\documentclass[11pt,twoside]{article}
\usepackage{asp2010}
\usepackage{amssymb}
\usepackage[latin1]{inputenc}

\resetcounters

\begin{document}

\title{HILTS, the Herschel Inspector and Long-Term Scheduler}
\author{Pedro G\'omez-Alvarez$^1$$^2$, Jon Brumfitt$^1$, Rosario Lorente$^1$ and Pedro Garc\'{\i}a-Lario$^1$
\affil{$^1$Herschel Science Centre, European Space Astronomy Centre (ESAC), European Space Agency (ESA), P.O. Box 78, Villanueva de la Cañada, 28691, Madrid, Spain}}
\affil{$^2$Ingenier\'{\i}a y Servicios Aeroespaciales, S.A. (INSA), Paseo Pintor Rosales 34, 28008, Madrid, Spain}

\begin{abstract}
Visualization, querying, statistical analysis and mid-long-term scheduling are common concerns for any observatory. HILTS is a Java tool developed for the Herschel project to address all these issues in a unified way.
\end{abstract}

\section{Introduction}
Herschel \citep{pilbratt2010} is an ESA cornerstone observatory mission launched on 14 May 2009. Herschel covers the range from 55 to 672 microns with three instruments: HIFI, PACS and SPIRE, see \citealp{O07_1_adassxx} in these proceedings.
If planning, visualization and inspection capabilities are important in any observatory, cryogenic space observatories such as Herschel, with a estimated lifetime of 3.5-4 years, call for additional efforts to maximize the observatory scientific return.

HILTS was initially conceived to assist Herschel medium and long-term planning. The tool is also useful to assess the mission's past, present and future status. Short-term mission planning for a given Operational Day (OD) is executed using the Herschel Scientific Mission Planning System (SMPS) \citep{brumfitt2005}, which generates satellite telecommands that are uplinked to Herschel on a daily basis. HILTS has been developed sharing the common object-oriented framework.

\section{The Mission}
Herschel operational database is populated by more than 50,000 observation requests pertaining to 1,300 proposals, from around 500 astronomers,  (see figure \ref{fig:statistics} for their geographic distribution). There are several factors impacting Herschel scheduling: helium optimization, slews minimization, proposal completion, scientific grades, Targets of Opportunity (ToOs), operational issues. Herschel has also thermal and communication constraints: the observatory attitude is constrained by the (anti)Sun, Earth, Moon and some planets, the observatory needs also to communicate with the ground station every 24 hours, during the so-called Daily Tele-Communication Period (DTCP).
\section{Tool description}
HILTS is a Java tool, whose main screen is divided into a set of panels (See figure \ref{fig:hilts}):
\begin{figure}[h!]
  	\centering
  	\includegraphics[scale=0.20]{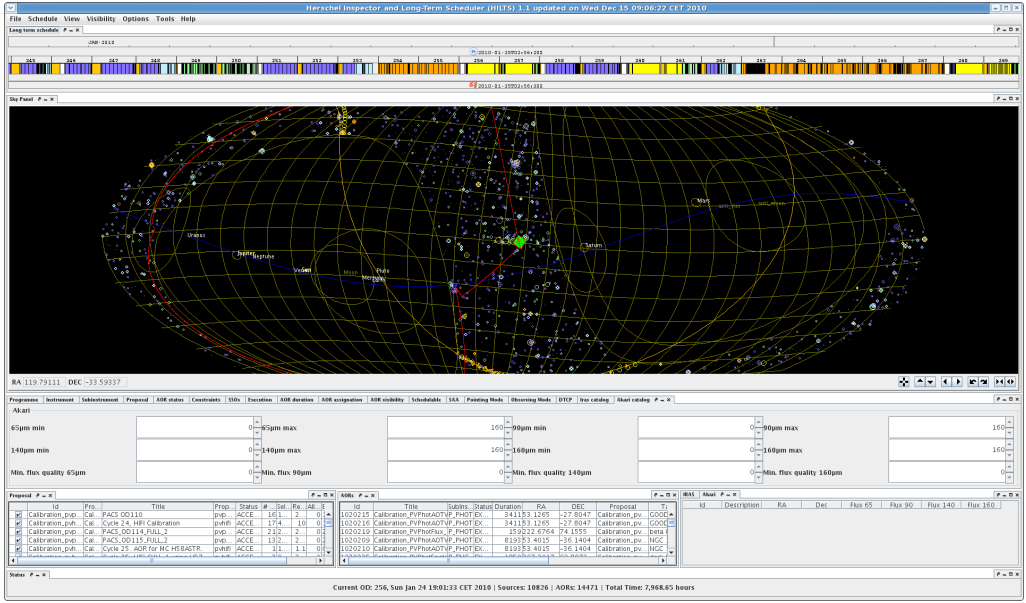}
  	\caption{HILTS main screen}
	\label{fig:hilts}
\end{figure}

\begin{itemize}
	\item \emph{Time panel:} It is composed of a set of horizontal sub-panels: A simple Gregorian calendar; a time selector, where the current time is selected; operational days, where the OD divisions are represented; scheduled observations, where the scheduled and already executed observations are represented; current observing block restrictions (groups of ODs preallocated to a given instrument mode) and the available scheduling interval.
	\item \emph{Sky panel:} Visible observations and current constraints are represented in this panel. The satellite pointing history for the current OD can also be plotted.
	\item \emph{Query panel:} Composed of multiple tabs, allowing arbitrary complex selections using the available criteria: Observation programmes, instruments and instrument modes, request status, Solar System Objects (SSOs), duration, etc.
	\item \emph{Proposal panel:} Current selected proposals are listed and can also be (de)selected.
	\item \emph{Requests panel:} Where the current selected observations are listed 
	\item \emph{Catalogs panel:} By default IRAS and AKARI catalogs. User catalogs can also be loaded. 
	\item \emph{Status panel:} General status information
\end{itemize}
All panels are interconnected with each other. For instance, when a new time is selected in the time panel, constraints and visible observations are updated simultaneously in the sky panel, while visible proposals and catalog objects are also updated in their respective panels. Selecting objects visible at a given time, is the default visibility selection. Other alternatives are available: (in)visibility during a time interval, during the DTCP, always visible, etc.

\subsection{Scheduling}
\begin{figure}[]
  	\centering
  	\includegraphics[scale=0.65]{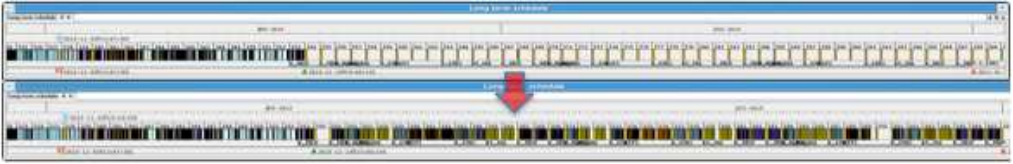}
  	\caption{Before and after an automatic scheduling run}
	\label{fig:scheduling}
\end{figure}

HILTS supports both manual and automatic scheduling. The former, by simple drag-and-drop from the observation panel to the scheduled observations sub-panel. The tool automatically places the observation at the earliest time within the dropped OD, taking into account observing blocks, observations duration, configuration and slews, amongst other factors. The latter (See figure \ref{fig:scheduling}) is attained by first selecting a suitable interval of typically several months and one of the set of pluggable strategies. For instance, if the ``remaining visibility'' strategy is selected, the tool will assign each of the visible observations by order of remaining visibility. HILTS scheduling is typically an iterative process: starting with one of the available ``filler'' strategies and finishing with an optimization phase (a simulated annealing optimization is being developed). Once a satisfactory schedule is obtained, it can be exported to XML and loaded into the SMPS.
\subsection{Statistics}
\begin{figure}[h!]
  	\centering
  	\includegraphics[scale=0.22]{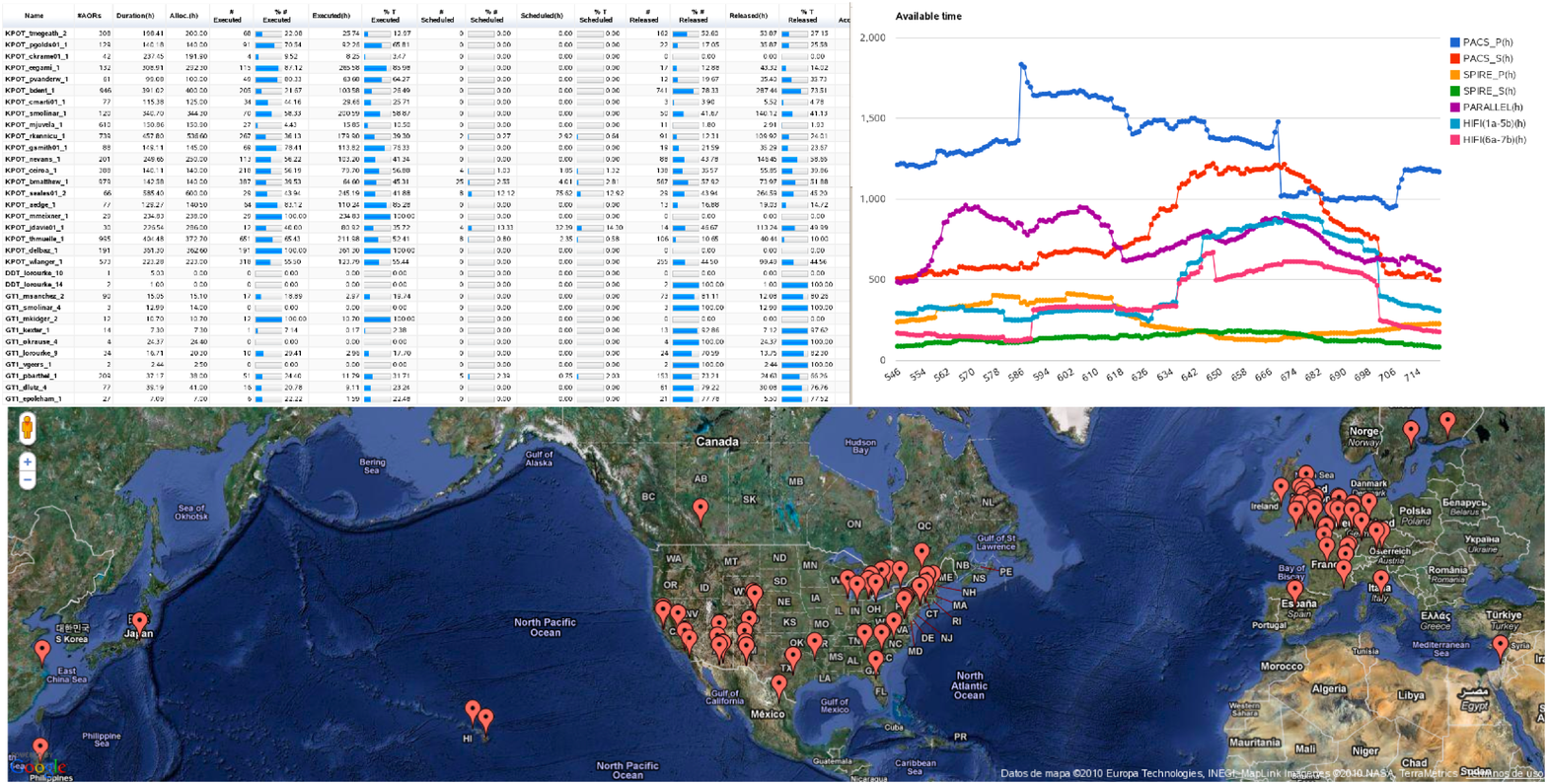}
  	\caption{Some examples of HILTS statistical capabilities: from upper left and clockwise, proposal completion report, available time per instrument mode during a long-term schedule and Herschel PIs geographical distribution.}
	\label{fig:statistics}
\end{figure}

HILTS can generate detailed statistics (See figure \ref{fig:statistics}) focused on several mission aspects, which help evaluating current mission status and thus retrofit observation strategies. Among the reports HILTS is capable to generate are: Execution reports where the completion of each programme, proposal and instrument mode is displayed; duplication studies where possible collisions between proposals are identified helping to avoid redundant science; scheduling reports, where several figures of merit and reports are generated as a result of a long-term schedule generation or an already executed period. The AJAX Google presentation API\footnote{http://code.google.com/apis/visualization/documentation/gallery.html} has been extensively used to implement this functionality.

\subsection{Catalogs and Virtual Observatory}
HILTS is also able to interact with on-line catalogs from Vizier \citep{vizier2006}. Specially relevant catalogs as IRAS and AKARI can be also filtered by flux in the query panel. A synthetic catalog of IR sources for the selection of candidate ``filler observations'' is also included.
\begin{figure}[h!]
  	\centering
  	\includegraphics[scale=0.22]{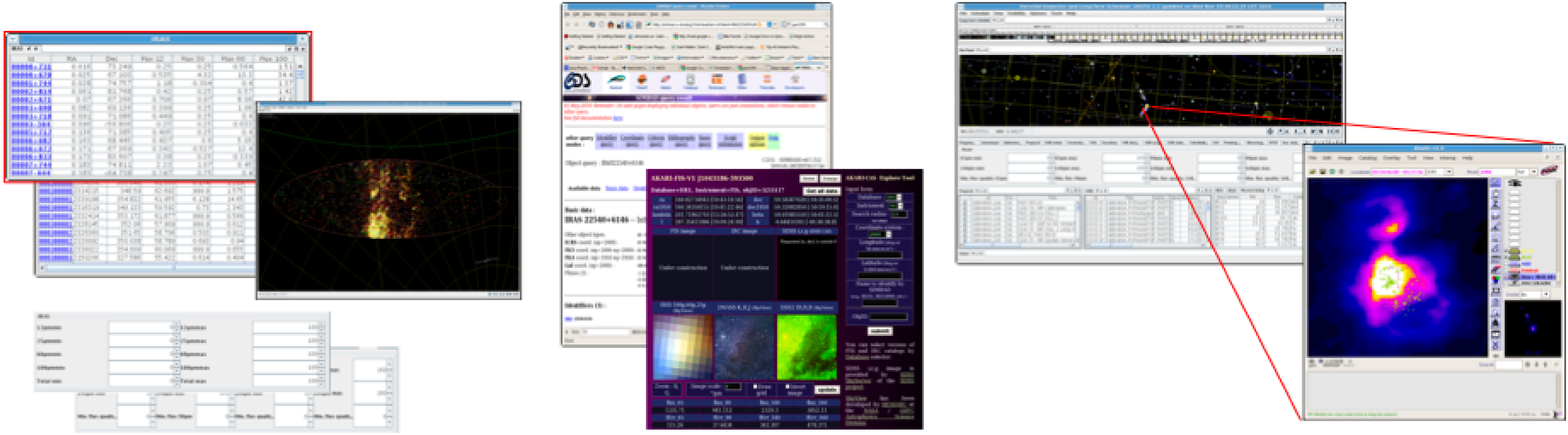}
  	\caption{Some screen-shots illustrating HILTS catalogs and VO interoperability: From left to right, IRAS sources available during DTCP, Vizier information of a given source and a joint HILTS-Aladin session centered at M42}
	\label{fig:cataloguesVO}
\end{figure}
HILTS can also inter-operate with VO tools as Aladin (See \citealp{F2_adassxx}) using the SAMP protocol (See figure \ref{fig:cataloguesVO}).

\bibliography{author}

\begin{thebibliography}{}
\expandafter\ifx\csname natexlab\endcsname\relax\def\natexlab#1{#1}\fi
\expandafter\ifx\csname url\endcsname\relax
  \def\url#1{\texttt{#1}}\fi
\expandafter\ifx\csname urlprefix\endcsname\relax\def\urlprefix{URL }\fi
\providecommand{\eprint}[2][]{\url{#2}}

\bibitem[{{Boch, Thomas}(2011)}]{F2_adassxx}
{Boch, Thomas} 2011, in ADASS XX, edited by I.~N. Evans, A.~Accomazzi, D.~J.
  Mink, \& A.~H. Rots, vol. TBD

\bibitem[{Brumfitt(2005)}]{brumfitt2005}
Brumfitt, J. 2005, in 5th Australian Space Science Conference

\bibitem[{{{Garc\'{\i}a-Lario}, P.G.}(2011)}]{O07_1_adassxx}
{{Garc\'{\i}a-Lario}, P.G.} 2011, in ADASS XX, edited by I.~N. Evans,
  A.~Accomazzi, D.~J. Mink, \& A.~H. Rots, vol. TBD

\bibitem[{{Genova} et~al.(2006){Genova}, {Allen}, {Bienayme}, {Boch},
  {Bonnarel}, {Cambresy}, {Derriere}, {Dubois}, {Fernique}, {Landais},
  {Lesteven}, {Loup}, {Oberto}, {Ochsenbein}, {Schaaff}, {Vollmer}, {Wenger},
  {Louys}, {Davoust}, \& {Jasniewicz}}]{vizier2006}
{Genova}, F., {Allen}, M.~G., {Bienayme}, O., {Boch}, T., {Bonnarel}, F.,
  {Cambresy}, L., {Derriere}, S., {Dubois}, P., {Fernique}, P., {Landais}, G.,
  {Lesteven}, S., {Loup}, C., {Oberto}, A., {Ochsenbein}, F., {Schaaff}, A.,
  {Vollmer}, B., {Wenger}, M., {Louys}, M., {Davoust}, E., \& {Jasniewicz}, G.
  2006, in Bulletin of the American Astronomical Society, vol.~38 of Bulletin
  of the American Astronomical Society, 1003

\bibitem[{{Pilbratt} et~al.(2010){Pilbratt}, {Riedinger}, {Passvogel}, {Crone},
  {Doyle}, {Gageur}, {Heras}, {Jewell}, {Metcalfe}, {Ott}, \&
  {Schmidt}}]{pilbratt2010}
{Pilbratt}, G.~L., {Riedinger}, J.~R., {Passvogel}, T., {Crone}, G., {Doyle},
  D., {Gageur}, U., {Heras}, A.~M., {Jewell}, C., {Metcalfe}, L., {Ott}, S., \&
  {Schmidt}, M. 2010, \aap, 518, L1+. \eprint{1005.5331}

\end{thebibliography}
\bibliographystyle{asp2010}

\end{document}